\documentclass[article]{JHEP3}

\usepackage{amsmath,amssymb}
\usepackage[dvips]{graphics}
\usepackage{epsfig}

\newcommand{\be}{\begin{equation}}
\newcommand{\ee}{\end{equation}}
\newcommand{\bea}{\begin{eqnarray}}
\newcommand{\eea}{\end{eqnarray}}
\newcommand{\bean}{\begin{eqnarray*}}
\newcommand{\eean}{\end{eqnarray*}}

\def\beq{\begin{equation}}

\def\eeq{\end{equation}}

\relax

\def\d{\partial}

\def\a{\alpha'}

\def\R{\mathcal{R}}

\def\MM{\boldsymbol{\mu}}

\preprint{
{\small{\textsf{CPHT-RR047.0805}}}
\\
{\small{\textsf{SPHT-T05/151}}}
\\
{\small{\textsf{ITFA-2006-19}}}
\\
{\small{\textsf{CERN-PH-TH/2006-076}}}
\\
{\small{\texttt{hep-th/0605001}}}
}

\title{Higher--Derivative Corrected Black Holes: Perturbative Stability and Absorption Cross--Section in Heterotic String Theory}

\author{Filipe Moura$^{a,b,c}$ and Ricardo Schiappa$^{d}$
\\
$^{a}$Centre de Physique Th\'eorique, \'Ecole Polyt\'echnique,\\
F91128 Palaiseau Cedex, France\\
\\
$^{b}$Service de Physique Th\'eorique, CEA/Saclay,\\
F91191 Gif--sur--Yvette Cedex, France\\
\\
$^{c}$Instituut voor Theoretische Fysica, Universiteit van Amsterdam,\\
Valckenierstraat 65, 1018 XE Amsterdam, The Netherlands\footnote{Present address.}\\
\\
$^{d}$Theory Division, Physics Department, CERN,\\
CH--1211 Gen\`eve 23, Switzerland\\
\\
\email{fmoura@spht.saclay.cea.fr}, \quad
\email{ricardos@mail.cern.ch}
}

\abstract{
This work addresses spherically symmetric, static black holes in higher--derivative stringy gravity. We focus on the curvature--squared correction to the Einstein--Hilbert action, present in both heterotic and bosonic string theory. The string theory low--energy effective action necessarily describes both a graviton and a dilaton, and we concentrate on the Callan--Myers--Perry solution in $d$--dimensions, describing stringy corrections to the Schwarzschild geometry. We develop the perturbation theory for the higher--derivative corrected action, along the guidelines of the Ishibashi--Kodama framework, focusing on tensor type gravitational perturbations. The potential obtained allows us to address the perturbative stability of the black hole solution, where we prove stability in any dimension. The equation describing gravitational perturbations to the Callan--Myers--Perry geometry also allows for a study of greybody factors and quasinormal frequencies. We address gravitational scattering at low frequencies, computing corrections arising from the curvature--squared term in the stringy action. We find that the absorption cross--section receives $\a$ corrections, even though it is still proportional to the area of the black hole event--horizon. We also suggest an expression for the absorption cross--section which could be valid to all orders in $\a$.
}

\keywords{Stringy Curvature Corrections, Black Holes in $d$--Dimensions, Stability, Hawking Emission}

\begin{document}



\vfill

\eject

\section{Introduction and Summary}

The scattering of gravitons in string theory is radically distinct from the scattering of gravitons within general relativity. This is not only clearly true at the quantum loop level (after all, general relativity is not a renormalizable theory, while string theory is expected to be finite), but is also true at the classical level: gravitational wave scattering in \textit{classical} string theory and in general relativity is totally different. An immediate consequence of this fact is that a Ricci flat manifold can \textit{not} be a solution to the string theory equations of motion \cite{gw86}; the evolution of graviton scattering must be described in some other way.

One obvious way is to remain within the string theoretic framework and compute scattering amplitudes using the standard stringy machinery. While this approach is certainly of great interest in a wide range of applications, it is perhaps not the most adequate one if one is interested in studying fully non--perturbative gravitational solutions, such as black holes. Another way to describe gravitational string physics is to include higher--order derivative corrections in the standard Einstein--Hilbert (EH) action and perturbatively construct a low--energy effective action describing the string physics. Of course that, in this approach, only an action including an infinite set of higher derivative corrections could fully reproduce graviton scattering in string theory. Nevertheless, a given low--energy effective action with higher derivative stringy corrections will reproduce string scattering amplitudes up to, and including, terms of the order of the momenta raised to the number of derivatives in the highest--order correction. In this set--up, the intrinsic string physics resides precisely in these higher derivative corrections and their determination is crucial if one wants to obtain a complete understanding of stringy phenomena. This is the framework in which we shall work.

The higher derivative corrections to the pure EH action naturally appear in diverse powers of the curvature tensor, as well as powers of its contractions. For instance, in the type II superstring the first correction appears as a combination of quartic powers of the curvature tensor \cite{gw86, gvz86a, gvz86b, gvz86c, gz86} and the resulting  low--energy effective action correctly reproduces stringy graviton scattering up to eighth order in the momenta. This naturally modifies the equations of motion and is thus simple to understand why Ricci flat manifolds may no longer be solutions. However, this also seems to raise a puzzle on what concerns superstring compactifications on Calabi--Yau (CY) manifolds \cite{gw86}. Indeed, CY spaces are Ricci flat. The point, of course, is that CY manifolds are also K\"ahler \cite{gw86}: the discussion above concerning scattering amplitudes no longer applies as there are no scattering processes described by K\"ahler manifolds (these manifolds cannot have Lorentzian signature due to the complex structure). Thus, Ricci--flat K\"ahler solutions seem to be in good shape to describe string compactification. The real story turns out to be a little more subtle: the equations of motion of type II string theory, including the quartic--curvature corrections, are actually \textit{not} satisfied by Ricci--flat K\"ahler metrics \cite{gvz86a, gvz86b, gvz86c}. The solution to this problem is that while the exact solution to the metric ends up not being Ricci flat, it is still the case that it can always be related to the initial Ricci flat metric by a non--local field redefinition\footnote{If one wishes to use sigma--model language, the field redefinitions of the metric are associated to the choice of subtraction scheme used to compute the beta--functions. It turns out that it is always possible to choose a subtraction scheme where Ricci--flat K\"ahler metrics will have vanishing beta--function to all orders in perturbation theory \cite{ns86}.} \cite{ns86, s86}. A detailed and recent discussion concerning some of these issues can be found in \cite{lps03}, where the authors have explicitly computed quartic--curvature corrections to the original Ricci--flat K\"ahler metric for non--compact CY spaces, choosing a particular subtraction scheme in which the CY manifold remains K\"ahler, but is no longer Ricci flat.

Our interest in the present work resides not in higher derivative corrections to string theory compactifications but in higher derivative corrections to string--theoretic black hole solutions. As such, we shall concentrate on the simpler quadratic curvature--corrections present in both heterotic and bosonic string theory \cite{ckp86, z85, chsw85}. One important aspect to have in mind when studying higher derivative corrections within a \textit{string} theoretic context, is the role of the dilaton field. In fact, in string theory, one can \textit{not} discard the dilaton when including higher derivative corrections to the low--energy effective action \cite{ckp86}: the curvature corrections act as a source in the dilaton field equations, so that if the Riemann tensor has any non--vanishing components so will the dilaton field be non--trivial throughout spacetime. In this way, solutions to the higher--derivative corrected field equations will necessarily describe \textit{both} graviton and dilaton fields. Another point to have in mind is the exact form of the quadratic--curvature correction \cite{z85}. While sigma--model considerations naturally yield the square of the Riemann tensor as the sole correction to include in the low--energy effective action \cite{ckp86, chsw85}, a Lagrangian which only includes a Riemann tensor squared correction will contain ghost particles in its spectrum \cite{z85}. This is really just a consequence of working with a low--energy effective action and one can simply resolve this annoyance with a field redefinition which amounts to replacing the Riemann tensor squared term in the effective action by the Gauss--Bonnet (GB) combination (also involving Ricci tensor squared and Ricci scalar squared) \cite{z85}. As our interest resides with black hole solutions, this is in fact a point we need not worry about.

Black hole solutions in curvature corrected gravity have a long history and we refer the reader to \cite{m98} for a review and a complete list of references. The effective theory we will focus upon in this work is the curvature--squared correction to the EH action, plus dilaton field \cite{ckp86, chsw85}, present in both heterotic and bosonic string theory, and the particular black hole solution we shall address is the Callan--Myers--Perry (CMP) solution to this low--energy effective action \cite{cmp89}, describing a static, spherically symmetric black hole. This solution describes stringy corrections to the well--known $d$--dimensional Schwarzschild solution (see, \textit{e.g.}, \cite{mp86}). The physics of the stringy corrections is quite interesting \cite{cmp89}: the strength of the dilaton coupling (\textit{i.e.}, the string coupling constant) decreases as one approaches the black hole, so that string interactions become weaker in the vicinity of the black hole. Furthermore, the black hole mass increases due to the higher--curvature corrections, while the black hole temperature decreases. It is particularly interesting to note that the expression for the temperature yields both maximum and zero temperature for definite values of the size of the black hole (the size of the event horizon) \cite{cmp89}. This could have profound implications in the understanding of the full Hawking evaporation process. As to the entropy of the black hole, it increases with respect to the original Schwarzschild result. It is important to notice that because the low--energy effective action is constructed perturbatively in higher--curvature corrections, also the solutions to this effective action must be found perturbatively in the specific dimensionless expansion parameter. In particular, any solutions one may find will \textit{not} be valid in regions of strong curvature and one cannot obtain stringy information concerning the resolution of singularities in general relativity. Still, there is a rich variety of phenomena which may be studied within this framework.

Let us be more concrete about the specific problems we wish to address in this paper. Having obtained a black hole solution to the curvature--corrected low--energy effective action, it is important to know whether this solution is stable or not. The analysis of the linear stability of four--dimensional black hole solutions in general relativity was first address a long time ago \cite{rw57, z70}. The procedure amounts to studying the linearised Einstein equations in the given background and proceeds with a decomposition of the gravitational perturbation in tensor spherical harmonics (for spherically symmetric backgrounds) in order to obtain a radial equation describing the propagation of linear perturbations. But it was not until recently that the black hole stability problem in general relativity was addressed within a $d$--dimensional setting \cite{gh02, ik03a, ik03b, ik03c, ik03d}. A set of equations describing linear gravitational perturbations to static, spherically symmetric black holes in any spacetime dimension $d>3$ was derived in \cite{ik03a, ik03c}. These equations are of Schr\"odinger type and will be denoted as the Ishibashi--Kodama (IK) master equations. The $d$--dimensional perturbations come in three types: tensor, vector and scalar type perturbations. This nomenclature refers to the tensorial behavior on the sphere ${\mathbb{S}}^{d-2}$ of each gauge--invariant type of perturbation. The IK master equations were used in \cite{ik03b, ik03c, ik03d} to study the stability of $d$--dimensional black holes in general relativity. In the present paper, we apply the IK framework to the stringy curvature--squared corrected action, focusing on tensor type gravitational perturbations. This is done in section 3. Tensor type perturbations only exist in dimension $d>4$ and they are the simplest of the three types of perturbations. We obtain an equation for the gauge--invariant perturbation which describes string theoretic corrections to the original IK master equation. This equation is still of Schr\"odinger type; the corrections appear as corrections to the tensor potential. We then use this equation to prove stability of the CMP black hole solution, in any spacetime dimension $d>4$. This is done in section 4. A full proof of stability would still require analysis of vector and scalar type perturbations. However, it was also argued in \cite{gh02} that it is expected that only tensor modes probe the ${\mathbb{S}}^{d-2}$ base manifold sufficiently enough in order to produce instabilities. Thus, our results point towards full stability of the stringy corrected black hole solution. We should further notice that the equation we obtain, describing gravitational perturbations, can be applied to any other static, spherically symmetric solutions of the curvature--corrected Einstein equations.

There has been some other work in the literature along analogous guidelines \cite{dg04, dg05a, dg05b, tm98}. However, these works have instead concentrated on GB corrections to the EH action; in particular they have focused upon the solutions described in \cite{bd85, bd86}. The solution in \cite{bd85} is non--dilatonic and is furthermore an exact solution to the action consisting of EH and GB terms. Lacking a dilaton and not taking into account the perturbative nature of the stringy low--energy effective action, the solution in \cite{bd85}, although interesting on its own, is thus not applicable in a string theory context. The stability of this solution was recently studied in \cite{dg04, dg05a, dg05b} using the IK framework. Studying tensor type gravitational perturbations, it was shown in these papers that static, spherically symmetric black hole solutions of the EH plus GB system are stable for $d>4$ and $d \not = 6$ \cite{dg04, dg05a}. The case of vector and scalar type gravitational perturbations turns out to be more complex, as while vector perturbations are stable, scalar perturbations do lead to instabilities of spherically symmetric EH--GB black holes \cite{dg05b}. It should be noted that the framework in \cite{dg04, dg05a, dg05b} is quite close to the one we deploy in the present paper, in spite of analyzing a distinct physical situation\footnote{Further studies on the stability of non--dilatonic solutions include, \textit{e.g.}, \cite{no02}.}. The solution described in \cite{bd86} studies the EH and GB system with the inclusion of the dilaton field. This action may be regarded as a [complicated] field redefinition of the action we use in the present paper. The authors construct a solution which has dilatonic charge (and is thus distinct, even under field redefinitions, of the CMP solution), but they fail to provide a full analytical expression to their final result. In spite of this, numerical work has been done \cite{tm98} studying the range of stability of this particular solution.

One further application of the perturbation equation we obtain in this paper concerns the calculation of greybody factors and quasinormal frequencies, in the CMP background geometry. This is required data in the study of Hawking emission spectra and quasinormal ringing, and could be of future interest as gravitational--wave astronomy becomes an experimental reality. In section 5 we address the investigation of gravitational scattering at low frequencies, obtaining corrections to the emission spectra due to the higher--order derivative corrections in the low--energy effective action. We use this result in order to compute the absorption cross--section at low--frequency which, in spite of the fact that it does receive $\a$ corrections, turns out to be proportional to the area of the black hole event--horizon, just like it is in the EH case \cite{dgm96}. The expression we obtain suggests a natural generalization which could be valid to all orders in $\a$.

\section{Higher Derivative Corrections in String Theory}

String theory low--energy effective actions have three different types of contributions, each with a different origin. There are classical terms, which come from the expansion in $\a$ (world--sheet loops). These represent the finite size of strings (as compared to point particles). Then there are the quantum terms, which  depend on the string coupling constant $g_s=\mbox{e}^{\phi}$, and these can either be perturbative (arising from spacetime loops) or non--perturbative. In this article we shall only consider the classical $\a$ corrections, neglecting any kind of stringy quantum corrections. For this to be possible, one must clearly have $g_s \ll 1$.

In this section we begin by reviewing the precise form of the leading higher--derivative $\a$ corrections to bosonic, heterotic and superstring theories. We will carefully write down the low--energy effective actions, and respective equations of motion, for both graviton and dilaton fields. Although these string theories also contain antisymmetric tensor fields in their massless spectra, these tensor fields can always be consistently set to zero. This shall be the case in the $\a$--corrected black hole solution we will use in this work. Also, although bosonic or supersymmetric string theory lives, respectively, in $26$ or in $10$ spacetime dimensions, in this article we shall always consider black hole spacetimes in generic $d$--dimensions. One can think of this as having compactified string theory on a flat torus, leaving uncompactified $d$ spacetime--dimensions.

A low--energy string theory effective action is written, in general, and in the string frame, as\footnote{Our conventions are $\Gamma^\sigma_{\mu\nu} = \frac{1}{2} g^{\sigma\rho} \left( \partial_\mu g_{\nu\rho} + \partial_\nu g_{\rho\mu} -
\partial_\rho g_{\mu\nu} \right)$ for the connection coefficients, ${\R^{\rho}}_{\sigma\mu\nu} = \partial_\mu \Gamma^\rho_{\nu\sigma} + \Gamma^\rho_{\mu\lambda} \Gamma^\lambda_{\nu\sigma} - \partial_\nu \Gamma^\rho_{\mu\sigma} - \Gamma^\rho_{\nu\lambda} \Gamma^\lambda_{\mu\sigma}$ for the Riemann tensor, $\R_{\mu\nu} = {\R^{\rho}}_{\mu\rho\nu}$ for the Ricci tensor, and $\R = {\R^{\mu}}_{\mu}$ for the Ricci scalar.}

\be \label{esf}
\frac{1}{2\kappa^2} \int \mbox{d}^dx \sqrt{-g}\ \mbox{e}^{-2 \phi} \Big( \R + 4 \left( \d^\mu \phi \right) \d_\mu \phi + z Y(\R) \Big) + \mbox{fermions}.
\ee 

\noindent
Here, $Y(\R)$ is a scalar polynomial in the Riemann tensor representing the higher derivative stringy corrections to the metric tensor field, and $z$ is, up to a numerical factor, the suitable power of the inverse string tension $\a$ for $Y(\R)$. The dilaton field is $\phi$, and we shall not make the fermionic terms explicit. The dilaton and graviton field equations which follow from the above effective action are, respectively,

\bea
\nabla^2 \phi - \left( \nabla \phi \right)^2 + \frac{1}{4} \R + \frac{1}{4} z Y(\R) &=& 0, \\ 
\R_{\mu\nu} + 2 \nabla_\mu \nabla_\nu \phi + z \frac{\delta Y(\R)}{\delta g^{\mu\nu}} &=& 0. 
\eea

For most purposes it is more convenient to write the stringy effective action (\ref{esf}) in the Einstein frame, rather than in the string frame (which is the frame arising naturally from sigma--model considerations). In order to achieve that, one needs to perform a redefinition of the metric by a conformal transformation involving the dilaton, which will also obviously affect the Riemann tensor \cite{gs87}:

\bea
g_{\mu\nu} &\rightarrow& \exp \left( \frac{4}{d-2} \phi \right) g_{\mu\nu}, \\
{\R_{\mu\nu}}^{\rho\sigma} &\rightarrow& {\widetilde{\R}_{\mu\nu}}^{\ \ \ \rho\sigma} = {\R_{\mu\nu}}^{\rho\sigma} - {\delta_{\left[\mu\right.}}^{\left[\rho\right.} \nabla_{\left.\nu \right]} \nabla^{\left.\sigma \right]} \phi.
\eea 

\noindent
The low--energy effective action thus reads, in the same order in $\a$ and up to total derivatives,

\be \label{eef}
\frac{1}{2\kappa^2} \int \mbox{d}^dx \sqrt{-g} \left( \R - \frac{4}{d-2} \left( \d^\mu \phi \right) \d_\mu \phi + z\ \mbox{e}^{\frac{4}{d-2} \left( 1 + w \right) \phi} Y(\widetilde{\R}) \right) + \mbox{fermions}.
\ee

\noindent
Here $w$ is the conformal weight of $Y(\R)$, with the convention that $w \left( g_{\mu\nu} \right) = +1$ and $w \left( g^{\mu\nu} \right) = -1$. The corrected equations of motion for the dilaton and graviton fields are, in the Einstein frame,

\bea
\nabla^2 \phi - \frac{z}{2}\ \mbox{e}^{\frac{4}{d-2} \left( 1 + w \right) \phi}\ Y(\R) &=& 0, \label{dfe} \\
\R_{\mu\nu} + z\ \mbox{e}^{\frac{4}{d-2} \left( 1 + w \right) \phi} \left( \frac{\delta Y(\R)}{\delta g^{\mu\nu}} + \frac{1}{d-2} Y(\R) g_{\mu\nu} - \frac{1}{d-2} g_{\mu\nu} g^{\rho\sigma} \frac{\delta Y(\R)}{\delta g^{\rho\sigma}} \right) &=& 0. \label{gfe}
\eea

\noindent
In both equations above we have eliminated certain terms involving derivatives of $\phi$, which would only contribute at higher orders in our perturbative parameter $z$.

\subsection{$\a$ Corrections to Heterotic and Bosonic Strings}

Both heterotic and bosonic string theories have higher derivative corrections already at first order in $\a$, and these corrections are at most quadratic in the Riemann tensor. For these theories, the gravitational correction $Y(\R)$ in (\ref{eef}) is given, to first order in $\a$, by $Y(\R) = \R^2 - 4 \R^{\mu\nu} \R_{\mu\nu} + \R^{\mu\nu\rho\sigma} \R_{\mu\nu\rho\sigma}$, the [four--dimensional] GB combination, in order to avoid ghosts as discussed in \cite{z85}. Since our focus will be on non--perturbative solutions of the classical theory, and not on quantization, we shall not be concerned with this issue. Furthermore, we are only considering a low--energy effective action which is perturbative in $\a$; in this way one can neglect the Ricci terms in $Y(\R)$, which from (\ref{gfe}) would only contribute at a higher order in $\a$, and take

\be
z\ Y(\R) \equiv \frac{\lambda}{2}\ \R^{\mu\nu\rho\sigma} \R_{\mu\nu\rho\sigma}, 
\ee 

\noindent
as in \cite{cmp89}. Here $w \left( Y(\R) \right) = -2$ and $\lambda = \frac{\a}{2}, \frac{\a}{4}$ and $0$, for bosonic, heterotic and type II strings, respectively. With this choice of derivative corrections, the corrected equations of motion for the dilaton and graviton fields are, to this order,

\bea
\nabla^2 \phi - \frac{\lambda}{4}\ \mbox{e}^{\frac{4}{2-d} \phi} \left( \R_{\rho\sigma\lambda\tau} \R^{\rho\sigma\lambda\tau} \right) &=& 0, \label{bdfe} \\
\R_{\mu\nu} + \lambda\ \mbox{e}^{\frac{4}{2-d} \phi} \left( \R_{\mu\rho\sigma\tau} {\R_{\nu}}^{\rho\sigma\tau} - \frac{1}{2(d-2)} g_{\mu\nu} \R_{\rho\sigma\lambda\tau} \R^{\rho\sigma\lambda\tau} \right) &=& 0. \label{bgfe}
\eea

\noindent
These are the field equations we shall focus upon, in the following.

\section{Gravitational Perturbations to the $\R^2$ Corrected Field Equations}

Having set--up the low--energy effective theory we wish to study, we proceed with the construction of the IK master equations for the resulting field equations. This will allow us to address stability, under gravitational perturbations, of the $\R^2$ corrected field equations. But first, let us begin with a comment of \cite{m98} concerning the general stability problem in higher--derivative effective actions. If one writes the higher--curvature field equations of motion as $\R_{\mu\nu} = \lambda J_{\mu\nu}$, with $J_{\mu\nu}$ some higher--derivative contribution and $\lambda$ the [dimensionfull] perturbative parameter 
associated to the higher curvature term in the action, one can then construct solutions perturbatively as $g_{\mu\nu} = g_{\mu\nu}^{(0)} + \lambda g_{\mu\nu}^{(1)} + {\mathcal{O}} (\lambda^{2})$. The perturbative equations of motion to solve follow as

\be
\R_{\mu\nu} [ g^{(0)} ] = 0 \quad \mbox{and} \quad
\square \big|_{g^{(0)}}\ g^{(1)}_{\mu\nu} = J_{\mu\nu} [ g^{(0)} ],
\ee

\noindent
where $\square \big|_{g^{(0)}}$ is the second--order differential operator that arises from the linearisation of the Ricci tensor about the unperturbed metric $g^{(0)}$. The first equation is simply the Einstein equation of motion, while the second amounts to solving for a linearised perturbation in the chosen background, including a source term. If one now considers a further perturbation to the corrected black hole background, one needs to be careful in organizing the new disturbance within the perturbative $\lambda$ expansion. One should expand as

\be
g_{\mu\nu} = \hat{g}_{\mu\nu} + h_{\mu\nu} = \hat{g}^{(0)}_{\mu\nu} + h^{(0)}_{\mu\nu} + \lambda \hat{g}^{(1)}_{\mu\nu} + \lambda h^{(1)}_{\mu\nu} + {\mathcal{O}} (\lambda^{2}),
\ee

\noindent
where $\hat{g}_{\mu\nu} = \hat{g}^{(0)}_{\mu\nu} + \lambda \hat{g}^{(1)}_{\mu\nu}$ is the background metric satisfying the previous equations of motion. The perturbation to the background metric will therefore satisfy

\be
\square \big|_{\hat{g}^{(0)}}\ h^{(0)}_{\mu\nu} = 0 \quad \mbox{and} \quad
\square \big|_{\hat{g}^{(0)}}\ h^{(1)}_{\mu\nu} = {\mathcal{J}}_{\mu\nu} [ \hat{g}^{(0)}, h^{(0)} ],
\ee

\noindent
where ${\mathcal{J}}_{\mu\nu} [ \hat{g}^{(0)}, h^{(0)} ]$ is the linearisation of $J_{\mu\nu} [ \hat{g}^{(0)} + h^{(0)} ]$. What is interesting to observe \cite{m98} is that perturbations to the background metric (and hence, the causal structure) are completely determined by the \textit{original} background metric. This also guarantees that the black hole solution does remain a black hole solution. The stability problem should now amount to analyzing whether the previous equations allow for runaway solutions. In the following we shall make these arguments fully precise, following \cite{ik03a, ik03c}.

\subsection{General Setup of the Perturbation Theory}

We are interested in studying the behavior, under gravitational perturbations, of string--corrected black hole solutions to the field equations discussed in the previous section. The analysis of these perturbations shall be carried out in a generic spacetime dimension $d$. As such, the most suitable tool to carry out such study is the $d$--dimensional perturbation theory framework developed by Ishibashi and Kodama, and which is itself a $d$--dimensional generalization of the original four--dimensional formalism of Regge and Wheeler \cite{rw57} and Zerilli \cite{z70}. The IK formalism was first set--up in \cite{iks00} and then further developed in a series of articles, where it was applied to the study of gravitational perturbations of maximally symmetric neutral \cite{ik03a} and charged \cite{ik03c} black holes in $d$--dimensions. As set--up, the IK framework applies to generic spacetimes of the form $\mathcal{M}^{d} = \mathcal{N}^{d-n} \times \mathcal{K}^n$, with coordinates $\left\{ x^\mu \right\} = \left\{ y^a, \theta^i \right\}$. In here $\mathcal{K}^n$ is a manifold with constant sectional curvature $K$, which describes the geometry of the black hole event horizon. The metric in the total space $\mathcal{M}^{d}$ is then written as

\be \label{ikmetric}
g = g_{ab}(y)\ dy^a \otimes dy^b + r^2(y)\ \gamma_{ij}(\theta)\ d\theta^i \otimes d \theta^j.
\ee

\noindent
For our purposes, we shall take $n = d-2$ and the $\mathcal{K}^n$ manifold, describing the geometry of the horizon, will be a $(d-2)$--sphere (thus, with $K=1$). Also, coordinates will be $\left\{ y^a \right\} = \left\{ t, r \right\}$ with $\left\{ r, \theta^i \right\}$ being the usual spherical coordinates so that $r(y) = r$ and $\gamma_{ij}(\theta)\ d\theta^i \otimes d\theta^j = d\Omega_{d-2}^2$.

One can consider perturbations either to the metric field or to any other physical field of the system under consideration. The most general perturbation will obviously involve perturbations to all the fields present in the low--energy effective action. Beginning with perturbations to the metric tensor field, these are given by 

\be
h_{\mu\nu} = \delta g_{\mu\nu} \quad \mbox{and} \quad h^{\mu\nu} = -\delta g^{\mu\nu}.
\ee

\noindent
From this variation one can easily compute the variation of the Riemann tensor, according to the Palatini equation

\be
\delta {\R^{\rho}}_{\sigma\mu\nu} = \nabla_\mu\ \delta \Gamma_{\nu\sigma}^\rho - \nabla_\nu\ \delta \Gamma_{\mu\sigma}^\rho.
\ee 

\noindent
Indeed, from

\be \label{deltagamma}
\delta \Gamma_{\mu\nu}^\rho = \frac{1}{2} \left( \nabla_\mu {h_{\nu}}^{\rho} + \nabla_\nu {h_{\mu}}^{\rho} - \nabla^{\rho} h_{\mu\nu} \right)
\ee 

\noindent
one immediately derives

\be \label{palatiniexp}
\delta \R_{\rho\sigma\mu\nu} = \frac{1}{2} \left( {\R_{\mu\nu\rho}}^{\lambda} h_{\lambda\sigma} - {\R_{\mu\nu\sigma}}^{\lambda} h_{\lambda \rho} - \nabla_\mu \nabla_\rho h_{\nu\sigma} + \nabla_\mu \nabla_\sigma h_{\nu\rho} - \nabla_\nu \nabla_\sigma h_{\mu\rho} + \nabla_\nu \nabla_\rho h_{\mu\sigma} \right).
\ee

General tensors, of rank at most equal to two, can be uniquely decomposed in tensor, vector and scalar components, according to their tensorial behavior on the $(d-2)$--sphere \cite{ik03a}. In particular, this is also true for the perturbations to the metric tensor field, and this is the whole basis for addressing perturbation theory within the IK framework \cite{iks00, ik03a} (it should be further noted that metric perturbations of tensor type only exist for dimensions $d>4$, unlike perturbations of vector and scalar type, which exist for dimension $d\ge4$). That this is a very convenient decomposition immediately follows from the fact that the geometry of the black hole event horizon is precisely that of a $(d-2)$--dimensional sphere.

In this work we shall only consider tensor type gravitational perturbations to the metric field, for $\a$--corrected $\R^2$ black holes in string theory (as we will show later, one can consistently set tensor type perturbations to the dilaton field to zero). These perturbations were worked out in detail in \cite{ik03a}, where it is shown that they can be written as

\be \label{htensor}
h_{ij} = 2 r^2 (y^a)\ H_T (y^a)\ \mathcal{T}_{ij} (\theta^i), \quad h_{ia} = 0, \quad h_{ab} = 0,
\ee 

\noindent
with $\mathcal{T}_{ij}$ satisfying

\be \label{propt}
\left( \gamma^{kl} D_k D_l + k_T \right) \mathcal{T}_{ij} = 0, \quad D^i \mathcal{T}_{ij} = 0, \quad g^{ij} \mathcal{T}_{ij} = 0.
\ee

\noindent
Here, $D_i$ is the covariant derivative on the $(d-2)$--dimensional sphere, associated with the metric $\gamma_{ij}$. Thus, the $\mathcal{T}_{ij}$ are the eigentensors of the Laplacian on the sphere ${\mathbb{S}}^{d-2}$, whose eigenvalues are given by $k_T + 2 = \ell \left( \ell + d - 3 \right)$, where $\ell = 2,3,4,\ldots$. It should be further noticed that the expansion coefficient $H_T$ is gauge--invariant by itself. This is rather important: when dealing with linear perturbations to a system with gauge invariance one might always worry that final results could be an artifact of the particular gauge one chooses to work with. Of course the simplest way out of this is to work with gauge--invariant variables, and this is precisely implemented in the IK framework \cite{iks00, ik03a, ik03c}. Observe that so far we have only made the choice of background metric we wish to perturb, and no choice of equations of motion plays any role. Thus, the IK gauge--invariant variables are also valid for higher derivative theories as long as diffeomorphisms keep implementing gauge transformations.

\subsection{Perturbation of a Spherically Symmetric Static Solution}

Let us now focus on a vacuum--type static, spherically symmetric background metric for the remainder of this section. Such a metric is clearly of the type (\ref{ikmetric}), and is given by\footnote{If one were to consider the most general static, spherically symmetric metric, then the $g_{rr}$ component of the metric would equal a generic function $g(r)$. It is the vacuum field equations which imply $g(r) = f^{-1}(r)$. This last condition is preserved by the solutions we shall considered in this article, even in the presence of the dilaton and higher--derivative corrections.}

\be \label{schwarz}
g = -f(r)\ dt \otimes dt + f^{-1}(r)\ dr \otimes dr + r^2 d\Omega^2_{d-2}.
\ee

\noindent
The nonzero components of the Riemann tensor for this metric are

\bea \label{ikriemann}
\R_{trtr} &=& \frac{1}{2} f''\,, \nonumber \\
\R_{itjt} &=& \frac{1}{2} \frac{ff'}{r} g_{ij}\,, \nonumber \\
\R_{irjr} &=& - \frac{1}{2} \frac{f'}{rf} g_{ij}\,, \nonumber \\\
\R_{ijkl} &=& \frac{1}{r^2} \big( 1 - f \big) \Big( g_{ik} g_{jl} - g_{il} g_{jk} \Big).
\eea

Perturbations to the Einstein field equations (without any $\a$ corrections) are described by an equation which is obtained by first perturbing the Ricci tensor $\mathcal{R}_{ij}$. If one collects the above expressions for $h_{\mu\nu}$ and the covariant derivatives, and replaces them on the Palatini equation presented above, one obtains

\bea
\delta \mathcal{R}_{ij} &=& \frac{r^2}{f} \left( \partial^2_t H_T \right) \mathcal{T}_{ij} - r^2 f \left( \partial^2_r H_T \right) \mathcal{T}_{ij} - r^2 f^\prime \left( \partial_r H_T \right) \mathcal{T}_{ij} - 2 r f^\prime H_T \mathcal{T}_{ij} + \left( 2 - d \right) r f \left( \partial_r H_T \right) \mathcal{T}_{ij} + \nonumber \\
&+& \left( 2 d - 4 \right) H_T \mathcal{T}_{ij} + \left( 6 - 2 d \right) f H_T \mathcal{T}_{ij} + k_T H_T \mathcal{T}_{ij} \,, \\
\delta \mathcal{R}_{ia} &=& 0, \quad 
\delta \mathcal{R}_{ab} = 0, \quad
\delta \mathcal{R} = 0.
\eea 

\noindent
By further using the so--called ``tortoise'' coordinate $x$, which is defined via $dx = \frac{dr}{f(r)}$, it was shown in \cite{ik03a} that the equation for the perturbations, $\delta \mathcal{R}_{ij} = 0$, can be written as a ``master equation'' for the gauge--invariant ``master variable'' $\Phi \equiv r^{\frac{d-2}{2}} H_T$. This master equation is of Schr\"odinger type:

\be \label{ikmaster}
- \frac{\partial^2 \Phi}{\partial x^2} + V_{\textsf{T}}(x) \Phi = - \frac{\partial^2 \Phi}{\partial t^2}. 
\ee

\noindent
Here, $V_{\textsf{T}}$ is the potential for tensor type gravitational perturbations, and is given in its most general form by \cite{ik03a}

\be \label{ikp}
V_{\textsf{T}} (r) = f(r) \left( \frac{\ell \left( \ell + d - 3 \right)}{r^2} + \frac{\left( d - 2 \right) \left( d - 4 \right) f(r)}{4r^2} + \frac{\left( d - 2 \right) f'(r)}{2r} \right). 
\ee

\noindent
Similar master equations of Schr\"odinger type can be obtained for the vector type and scalar type gravitational perturbations. However, their respective potentials are more complicated \cite{ik03a, ik03c}. Let us stress that the above master equation is valid for any static, spherically symmetric solution to the Einstein field equations of the form (\ref{schwarz}), but so far \textit{without} any $\a$ corrections. Our goal in the following is to determine which kind of master equation one will obtain when considering perturbations to the solutions of the $\a$--corrected field equations. It is expected that the potential above will get corrected.

In order to determine an equation for $H_T$ with inclusion of the $\a$ corrections from (\ref{eef}), one needs to perturb the $\a$--corrected field equations (\ref{bdfe}) and (\ref{bgfe}). In order to do so, and for both of them, one first needs to obtain the variation of the Riemann tensor under generic perturbations of the metric, which is given by (\ref{palatiniexp}). Replacing $h_{\mu\nu}$ by the expressions given in (\ref{htensor}) and further using the expressions for the components of the Riemann tensor as in (\ref{ikriemann}), one obtains 

\bea
\delta\R_{ijkl} &=& \Big( \big( 2 f - 1 \big) H_T + f \partial_r H_T \Big) \Big( g_{il} \mathcal{T}_{jk} - g_{ik} \mathcal{T}_{jl} - g_{jl} \mathcal{T}_{ik} + g_{jk} \mathcal{T}_{il} \Big) + \nonumber \\
&+& r^2 H_T \Big( D_i D_l \mathcal{T}_{jk} - D_i D_k \mathcal{T}_{jl} - D_j D_l \mathcal{T}_{ik} + D_j D_k \mathcal{T}_{il} \Big), \label{drtensori} \\
\delta\R_{itjt} &=& \left( - r^2 \partial_t^2 H_T + \frac{1}{2} r^2 f f' \partial_r H_T + r f f' H_T \right) \mathcal{T}_{ij}\,, \\
\delta\R_{irjr} &=& \left( - r \frac{f'}{f} H_T - \frac{1}{2} r^2 \frac{f'}{f} \partial_r H_T - 2 r \partial_r H_T - r^2 \partial^2_r H_T \right) \mathcal{T}_{ij}\,, \\
\delta\R_{abcd} &=& 0. \label{drtensora}
\eea

\noindent
By perturbing (\ref{bdfe}) and (\ref{bgfe}) one further gets

\bea
\delta \nabla^2 \phi &-& \frac{\lambda}{4}\ \mbox{e}^{\frac{4}{2-d} \phi}\ \delta \left( \R_{\rho\sigma\lambda\tau} \R^{\rho\sigma\lambda\tau} \right) + \frac{\lambda}{d-2}\ \mbox{e}^{\frac{4}{2-d} \phi}\ \R_{\rho\sigma\lambda\tau} \R^{\rho\sigma\lambda\tau}\ \delta \phi = 0, \label{pbdfe} \\
\delta \R_{ij} &+& \lambda\ \mbox{e}^{\frac{4}{2-d} \phi} \left[ \delta \left( \R_{i\rho\sigma\tau} {\R_{j}}^{\rho\sigma\tau} \right) - \frac{1}{2(d-2)} \R_{\rho\sigma\lambda\tau} \R^{\rho\sigma\lambda\tau}\ h_{ij} - \right. \nonumber \\
&-& \left. \frac{1}{2(d-2)}\ g_{ij}\ \delta \left( \R_{\rho\sigma\lambda\tau} \R^{\rho\sigma\lambda\tau} \right) \right] + \frac{4}{d-2}\ \R_{ij}\ \delta \phi = 0. \label{pbgfe}
\eea 

\noindent
Using the explicit form of the Riemann tensor (\ref{ikriemann}), alongside with the variations (\ref{htensor}) and (\ref{drtensori}--\ref{drtensora}), one can straightforwardly compute most of the terms in (\ref{pbdfe}) and (\ref{pbgfe}). What remains to be analyzed is the equation describing the dilaton perturbation, $\delta \phi$. In this framework, we have

\be
\delta \nabla^2 \phi = g^{ab} \delta \left( \nabla_a \nabla_b \phi \right) - h^{ij} \nabla_i \nabla_j \phi + g^{ij} \delta \left( \nabla_i \nabla_j \phi \right).
\ee

\noindent
From (\ref{deltagamma}) one can easily show that both $\delta \Gamma_{ab}^{c} = 0$ and $g^{ij} \delta \Gamma_{ij}^{a} = 0$. Assuming (as it is the present case, due to the spherical symmetry of the background) that the dilaton does not depend on the angular coordinates, \textit{i.e.}, that $\partial_k \phi = 0$, we are left with the following

\be
\delta \nabla^2 \phi = g^{ab} \partial_a \partial_b \delta \phi - g^{ab} \Gamma_{ab}^{c} \partial_c \delta \phi + g^{ij} \partial_i \partial_j \delta \phi - g^{ij} \Gamma_{ij}^{k} \partial_k \delta \phi - g^{ij} \Gamma_{ij}^{a} \partial_a \delta \phi.
\ee

\noindent
Using this result in (\ref{pbdfe}), and after showing that it turns out that $\delta \left( \R_{\rho\sigma\lambda\tau} \R^{\rho\sigma\lambda\tau} \right) = 0$ in our situation, one concludes that the equation describing perturbations to the dilaton field is a homogeneous differential equation for $\delta \phi$. Thus, there is no immediate inconsistency in setting $\delta \phi = 0$. Because furthermore the dilaton is a scalar field, which does not admit tensor type perturbations, we are led to conclude that it is indeed the case that $\delta \phi = 0$\footnote{Let us point out that our derivation of this result is only valid for tensor type perturbations to the action we are considering. Let us also point out that we do not expect this to hold for other types of gravitational perturbations, in particular this will most certainly not happen for scalar type perturbations. Rather, one will find coupled equations between perturbations of the metric field and of the dilaton field.}. This is the information one obtains from (\ref{pbdfe}).

Collecting the several expressions, the result for (\ref{pbgfe}) finally becomes

\bea
&&
\left[ 1 - 2 \lambda \frac{f'(r)}{r} \right] \Big( \partial^2_t H_T - f^2 (r)\,\partial^2_r H_T \Big) - \nonumber \\
&&
- f(r) \left[ ( d - 2 ) \frac{f(r)}{r} + f'(r) + \frac{2 \lambda}{r} \left( 2 ( d - 4 ) \frac{f(r) \left( 1 - f(r) \right)}{r^2} - 2 \frac{f(r) f'(r)}{r} - \left( f'(r) \right)^2 \right) \right] \partial_r H_T + \nonumber \\
&&
+ \frac{f(r)}{r} \left[ \frac{\ell \left( \ell + d - 3 \right)}{r} - 2 f'(r) 
+ 2 ( d - 3 ) \frac{1 - f(r)}{r} \right. + \nonumber \\ 
&&
+ \left. \frac{\lambda}{r} \left( 4 \ell \left( \ell + d - 3 \right) \frac{1 - f(r)}{r^2} + 2 \left( d - 3 \right)\frac{\left( 1 - f(r) \right)^2}{r^2} - r^2 \frac{\left( f''(r) \right)^2}{d-2} \right) \right] H_T = 0. \label{master0} 
\eea

\noindent
As it stands, the above equation for the perturbations looks rather complicated, specially if compared to (\ref{ikmaster}) and (\ref{ikp}). Indeed, this expression does not even have the Schr\"odinger form usually associated to the master equations for gravitational perturbations. So, we would now like to re--write the equation above in the form of a Schr\"odinger--like master equation, as in (\ref{ikmaster}). In order to achieve so, one first writes the perturbation equation in terms of the tortoise coordinate, $x$. Then, and following a procedure rather similar to the one in \cite{dg05a}, one first considers a general equation of the form

\be \label{dottigen}
\partial^2_t H_T - F^2(r)\ \partial^2_r H_T + P(r)\ \partial_r H_T + Q(r)\ H_T = 0.
\ee

\noindent
Further defining

\be
k(r) = \frac{1}{\sqrt{F}} \exp \left( - \int dr\ \frac{P}{2F^2} \right), \quad \Phi = k(r) H_T,
\ee

\noindent
it is easy to see that (\ref{dottigen}) may then be written as a Schr\"odinger--type equation,

\be
\frac{\partial^2 \Phi}{\partial x^2} - \frac{\partial^2 \Phi }{\partial t^2} = \left( Q + \frac{F'^2}{4} - \frac{F F''}{2} - \frac{P'}{2} + \frac{P^2}{4 F^2} + \frac{P F'}{F} \right) \Phi \equiv V \left[ f(r) \right] \Phi, \label{potential0}
\ee 

\noindent
where $\Phi$ is the gauge--invariant ``master variable'' for the gravitational perturbation in the present framework, including $\a$ corrections. In our case, from (\ref{master0}) it immediately follows

\bea
F &=& f, \nonumber \\
P &=& - \frac{f}{1 - 2 \lambda \frac{f'}{r}} \left[ (d-2) \frac{f}{r} + f' + \frac{2 \lambda}{r} \left( 2 ( d - 4 ) \frac{f \left( 1 - f \right)}{r^2} 
- 2 \frac{ff'}{r} - f'^2  \right) \right], \nonumber \\ 
Q &=& \frac{f}{1 - 2 \lambda \frac{f'}{r}} \left[ \frac{\ell \left( \ell + d - 3 \right)}{r^2} - \frac{2 f'}{r} + 2 (d-3) \frac{1-f}{r^2} + \right. \nonumber \\ &+& \left. \frac{\lambda}{r^2} \left( 4 \ell \left( \ell + d - 3 \right) \frac{1-f}{r^2} + 2 (d-3) \frac{\left( 1-f \right)^2}{r^2} - \frac{\left( r f'' \right)^2}{d-2} \right) \right]. \label{fpqnp}
\eea

\noindent
While the algebraic manipulations above have produced non--polynomial terms in $\lambda$, it is important to recall the physics of the problem at hand. Indeed, in the present context, any black hole solution is built perturbatively in $\lambda$ and a solution, characterized by a function $f(r)$, will only be valid in regions where $r \gg \sqrt{\lambda}$, \textit{i.e.}, any perturbative solution is only valid for black holes whose event horizon is much bigger than the string length. Working to first order in the perturbative parameter $\lambda$ throughout then requires the above formulae to obey the same criteria. A simple power series expansion yields

\bea
F &=& f, \nonumber \\
P &=& - f \left[ (d-2) \frac{f}{r} + f' + 2 \lambda (d-4) \frac{f}{r^2} \left( 2  \frac{1 - f}{r} + f' \right) \right], \nonumber \\ 
Q &=& f \left[ \frac{\ell \left( \ell + d - 3 \right)}{r^2} - \frac{2 f'}{r} + 2 (d-3) \frac{1-f}{r^2} + \right. \nonumber \\ &+& \left. \frac{\lambda}{r^2} \left( \frac{2 \ell \left( \ell + d - 3 \right)}{r} \left( 2 \frac{1-f}{r} + f' \right) + \frac{2 (d-3) \left( 1-f \right)}{r} \left( \frac{1-f}{r} + 2f' \right) - 4 \left( f' \right)^2 - \frac{\left( r f'' \right)^2}{d-2} \right) \right]. \label{fpq}
\eea

\noindent
From (\ref{potential0}) and (\ref{fpq}) one finally obtains

\bea
V_{\textsf{T}} [f(r)] &=& f(r) \left( \frac{\ell \left( \ell + d - 3 \right)}{r^2} + \frac{\left( d - 2 \right) \left( d - 4 \right) f(r)}{4r^2} + \frac{2 \left( d - 3 \right) \left( 1 - f(r) \right)}{r^2} + \frac{\left( d - 6 \right) f'(r)}{2r} \right) + \nonumber \\
&+& \lambda\ \frac{f(r)}{r^2} \left[ \left( \frac{2 \ell \left( \ell + d - 3 \right)}{r} + \frac{\left( d - 4 \right) \left( d - 5 \right) f(r)}{r} + \frac{\left( d - 3 \right) \left( 1 - f(r) \right)}{r} + \left( d - 4 \right) f'(r) \right) \right. \times \nonumber \\
&\times& \left( 2 \frac{1 - f(r)}{r} + f'(r) \right) + \Big( 3 (d-3) - (4d-13) f(r) \Big) \frac{f'(r)}{r} - \nonumber \\
&-& \left. 4 \left( f'(r) \right)^2 + \left( d-4 \right) f(r) f''(r) - \frac{\left( r f''(r) \right)^2}{d-2} \right].
\label{potential}
\eea

\noindent
This is the potential for tensor--type gravitational perturbations of any kind of static, spherically symmetric $\R^2$ string--corrected black hole in $d$--dimensions. This is also one of the main results in this paper.

\subsection{On the Proof of Perturbative Stability}

In order to study the stability of a solution, we shall use the so--called ``S--deformation approach'', first introduced in \cite{ik03b} and later further developed in \cite{dg04, dg05a, dg05b}. Let us briefly review this technique in the following (for more details we refer the reader to the original discussion in \cite{ik03b}). After having obtained the potential (\ref{potential}) for the master equation (\ref{ikmaster}), one assumes that its solutions are of the form $\Phi(x,t) = e^{i\omega t} \phi(x)$, such that $\frac{\partial\Phi}{\partial t} = i\omega \Phi$. In this way the master equation may be written in Schr\"odinger form, as

\be
\left[ - \frac{d^2}{dx^2} + V(x) \right] \phi(x) \equiv A \phi(x) = \omega^2
\phi(x).
\ee

\noindent
A given solution of the gravitational field equations will then be perturbatively stable if and only if the operator $A$, defined above, has no negative eigenvalues for $x \in {\mathbb{R}}$ \cite{ik03b}. Technically, $A$ needs to be a positive self--adjoint operator in the Hilbert space of square--integrable functions of $x$. Considering a set of smooth functions $\{ \phi(x) \}$, on the range of $x$, the above condition is equivalent to the positivity---for any given $\phi$---of the following inner product \cite{ik03b}

\be
\left \langle \phi \left| A \phi \right \rangle \right. = \int_{-\infty}^{+\infty} \phi^\dagger (x) \left[ - \frac{d^2}{dx^2} + V(x) \right] \phi(x)\ dx = \int_{-\infty}^{+\infty} \left[ \left| \frac{d\phi}{dx} \right|^2 + V(x) \left|\phi \right|^2 \right] dx,
\ee

\noindent
where, in the last step, we have made an integration by parts. After some further algebra, and another integration by parts, one finds that the inner product $\left \langle \phi \left| A \phi \right \rangle \right.$ may be re--written as

\be
\left \langle \phi \left| A \phi \right \rangle \right. = \int_{-\infty}^{+\infty} \left[ \left| D \phi \right|^2 + \widetilde{V}(x) \left| \phi \right|^2 \right] dx,
\ee

\noindent
where we have defined $D = \frac{d}{dx} + S$ and $\widetilde{V}(x) = V(x) + f \frac{dS}{dr} - S^2$, with $S$ a completely arbitrary function. The stability is then guaranteed by the positivity of $\widetilde{V}(x)$, for $x \in {\mathbb{R}}$, whatever function $S$ is chosen \cite{ik03b}. If one now follows \cite{dg04}, and takes into consideration the form of the potential $V(x)$ given in (\ref{potential0}), the most convenient choice for $S$, \textit{i.e.}, the one which leads to the simplest analysis, is

\be
S = - \frac{F}{k} \frac{dk}{dr}.
\ee

\noindent
Indeed, with this choice of $S$, we simply obtain $\widetilde{V}(x) = Q$ and are left with

\be
\left \langle \phi \left| A \phi \right \rangle \right. = \int_{-\infty}^{+\infty} \left|D \phi \right|^2 dx + \int_{-\infty}^{+\infty} Q(x) \left|\phi \right|^2 dx.
\ee

\noindent
Thus, all that is necessary in order to guarantee the stability of a given solution is to check the positivity of $Q$, for $x \in {\mathbb{R}}$, with $Q$  given by (\ref{fpq}). This is particularly interesting if one recalls the definition of the tortoise coordinate, $dx = \frac{dr}{f(r)}$. It is simple to realize that the second term of the expression above becomes

\be
\int_{R_H}^{+\infty} Q(r) \frac{\left|\phi \right|^2}{f(r)} dr,
\ee

\noindent
where $R_H$ is the radius of the black hole event horizon. Because $f(r)$ is a positive function for $r > R_H$, perturbative stability of a given black hole solution, with respect to tensor--type gravitational perturbations, then follows if and only if one can prove that $Q(r)$ is a positive function for $r \ge R_H$, \textit{i.e.}, is a positive function outside of the black hole event horizon. We shall come back to this question in the following.

\section{The Callan--Myers--Perry Black Hole Solution}

The perturbation theory we have developed in the previous section is valid for any static, spherically symmetric solution to the field equations (\ref{bdfe}) and (\ref{bgfe}), \textit{i.e.}, for any background metric of the type (\ref{schwarz}). For the remainder of this work we shall focus on a particular black hole solution and study both perturbation and scattering theories associated to this particular solution, within the framework we have developed. But first, let us make some generic remarks concerning black hole solutions in string theory.

As noticed in \cite{ms93, kmrtw96, kt97}, the presence of higher--order curvature terms in the low--energy effective action affects the energy--momentum tensor in such a way that its time component, ${T_t}^t$, representing the local energy--density in the Einstein case, will not be necessarily positive--definite anymore. It so happens that this assumption, of positive definiteness, is a necessary condition for the no--hair conjecture. Another necessary condition for this conjecture, which again is not necessarily verified by the $\a$--corrected solutions, is the relation ${T_t}^t = {T_\theta}^\theta$ between time and angular components of the energy--momentum tensor (a relation which was checked to be valid in the case of spherically symmetric solutions to the Einstein theory). As these conditions are no longer verified for higher--derivative corrected black hole solutions, it may thus be possible to circumvent the no--hair conjecture and find several static, spherically symmetric black hole solutions to the curvature--corrected field equations. These new black hole solutions may have either primary or secondary hair outside the horizon. In the first case, of primary hair, besides introducing new fields outside of the horizon, the solutions also introduce new parameters (\textit{e.g.}, dilatonic charges). In the second case, of secondary hair, there are no new parameters and only new fields are introduced, the charges of which being eventually expressed in terms of the original parameters.

One of the new fields which must always be present in any string theoretic black hole solution is the dilaton field. As we have said before, a constant dilaton solution is only possible at the classical level, without any stringy corrections, as the curvature--correction terms act as a source for the dilaton field equation. Since we work in a string theoretic framework one could also expect that these black hole solutions should include, besides the graviton and the dilaton which we have alluded to above, couplings to different antisymmetric tensor fields (depending on the bosonic massless spectrum of each theory). But, in this case, it turns out that one can always consistently set the antisymmetric tensor fields to zero, without restricting the dilaton and graviton terms through the field equations. The same holds true for any fermionic fields arising in heterotic string theory. Still, considering black hole solutions with both higher derivative corrections and couplings to antisymmetric tensor fields would be interesting on its own, but, to our knowledge, these solutions have only been studied in four dimensions, in \cite{ms93, kmrtw96, kt97}.

Solutions with dilatonic primary hair (\textit{i.e.}, an independent dilatonic charge) were studied in \cite{bd86}. We shall instead consider the case of secondary hair. The black hole solution we will focus upon in this paper is a $d$--dimensional $\R^2$ stringy--correction to the Schwarzschild solution, and is thus naturally of the type of (\ref{schwarz}). This higher--derivative corrected solution was first obtained by Callan, Myers and Perry in \cite{cmp89} and is the mentioned CMP black hole solution. Its only free background parameter is $\MM$, a parameter which is related to the classical ADM mass of the black hole through

\be
M_{\mathrm{cl}} = \frac{\left( d-2 \right) A_{d-2}}{\kappa^2} \MM,
\ee

\noindent
where $A_n$ is the area of the unit $n$--sphere. For the classical $d$--dimensional Schwarzschild solution, we have

\be
f(r) = 1 - \frac{2\MM}{r^{d-3}}.
\ee

\noindent
In order to introduce $\a$ corrections to this solution, one chooses a coordinate system in which the position of the horizon, given by $r = \left( 2 \MM \right)^{\frac{1}{d-3}} \equiv R_H$ in the solution above, is not changed. Such a solution was worked out in \cite{cmp89}, perturbatively in $\lambda$, and in such a coordinate system $f(r)$ is given by

\be \label{fr2}
f(r) = \left( 1 - \frac{R_H^{d-3}}{r^{d-3}} \right) \left( 1 - \frac{(d-3)(d-4)}{2}\ \frac{\lambda}{R_H^2}\ \frac{R^{d-3}_H}{r^{d-3}}\ \frac{1 - \frac{R_H^{d-1}}{r^{d-1}}}{1 - \frac{R^{d-3}_H}{r^{d-3}}} \right).
\ee

\noindent
The $\a$--corrected ADM mass for this stringy black hole is given by

\be
M = \frac{\left( d-2 \right) A_{d-2}}{2\kappa^2} \lim_{r \to \infty} r^{d-3} \Big( 1 - f \left(r\right) \Big) = \left( 1 + \frac{(d-3)(d-4)}{2}\ \frac{\lambda}{R_H^2} \right) \frac{\left(d-2\right) A_{d-2}}{\kappa^2} \MM.
\ee

\noindent
This represents an ADM mass which is both inertial \textit{and} gravitational, since in the CMP solution we are considering, and to first order in $\lambda$, $ - g_{tt} = \frac{1}{g_{rr}} = f(r)$. The dilaton, which is classically set to vanish, will get $\a$--corrections. This means it will be of the form $\phi(r) = \lambda \varphi(r)$, with $\varphi(r)$ a solution to the dilaton equations of motion without any explicit dependence on the perturbative parameter $\lambda$. One further realizes that the dilaton corrections will only affect the field equations at an order in $\lambda$ which is higher than the one we are considering, and therefore we will not need it. Its full explicit expression may be found in \cite{cmp89}.

In order to compute the black hole temperature \cite{y84}, one first Wick--rotates to Euclidean time $t = i \tau$. The resulting manifold is smooth as long as $\tau$ is a periodic variable, with a period $\beta$ which is related to the black hole temperature as $T = \frac{1}{\beta}$. The precise smoothness condition amounts to

\be
2 \pi = \lim_{r \to R_H} \frac{\beta}{f^{-\frac{1}{2}} \left(r\right)} \frac{d f^{\frac{1}{2}}\left(r\right)}{d r},
\ee

\noindent
from which follows $T=\frac{1}{4 \pi} f' \left( R_H \right)$. In our particular case, this immediately implies

\be
T = \frac{d-3}{4 \pi R_H} \left( 1 - \frac{\left( d-1 \right) \left( d-4 \right)}{2}\ \frac{\lambda}{R_H^2} \right).
\ee

\noindent
One learns that the stringy $\a$--corrections actually \textit{reduce} the black hole temperature for $d>4$. This seems to suggest that black hole temperature may reach a maximum value. Indeed, taking the above formula literally, one finds, in the case of the bosonic string, a maximal temperature of $T_{\mathrm{max}} \sim \frac{0.06}{\sqrt{\alpha'}}$, with $d$ in the range $d=5$ to $d=26$, for finite black hole radius \cite{cmp89}. Similarly, one finds, in the case of the heterotic string, a maximal temperature of $T_{\mathrm{max}} \sim \frac{0.10}{\sqrt{\alpha'}}$, with $d$ in the range $d=5$ to= $d=10$, again for finite black hole radius \cite{cmp89}. Unfortunately, in both cases, the black hole radius at which this happens is of the order of the string scale and thus outside the validity range of the perturbative solution. Nevertheless, one may notice that this value of the maximal temperature is close to the Hagedorn temperature in the free string spectrum, $T_{\mathrm{critical}} \simeq \frac{0.08}{\sqrt{\alpha'}}$ for bosonic strings and $T_{\mathrm{critical}} \simeq \frac{0.16}{\sqrt{\alpha'}}$ for heterotic strings. Furthermore, the temperature formula yields zero temperature at a radius smaller than the one for maximal temperature \cite{cmp89}. Such a black hole could presumably be regarded as a string soliton. Let us finally add that if one computes the entropy of these stringy black holes, by a simple integration of the first law, it follows that

\be
S= \frac{A_{H}}{4} \left( 1 + (d-2)(d-3)\ \frac{\lambda}{R_H^2} \right),
\ee

\noindent
so that entropy no longer equals one quarter of the horizon area \cite{cmp89}. Indeed, the entropy of black holes is increased in string theory according to Wald's formula (see, \textit{e.g.}, \cite{w93, jkm93, n02, kl05}).

Having reviewed the black hole solution which we have decided to focus upon, in the following we shall analyze its stability in detail, using the machinery we have developed earlier. Then, in section 5, we shall further analyze scattering in this spacetime geometry.

\subsection{Potential for Classically Non--Dilatonic Solutions}

As described above, the CMP solution has vanishing dilaton field at the classical level (\textit{i.e.}, at order ${\mathcal{O}} (\lambda^0)$), while at first order in $\lambda$ it has a non--trivial dilaton field. It turns out that in this case one may further simplify some of our previous results concerning the perturbation theory of spherically--symmetric static solutions. Indeed, upon the further assumption that the dilaton field is of the type $\phi (r) = \lambda \varphi (r)$, one may judiciously use the field equation for $\R_{ij}$ in (\ref{bgfe}) to first order in $\lambda$ (\textit{i.e.}, neglecting all the dilaton terms, which would only contribute at least to order ${\mathcal{O}} (\lambda^2)$), and derive the relation

\be \label{f''}
\lambda \R_{abcd} \R^{abcd} = \lambda \left( f'' \right)^2 = 2 g^{ij} \R_{ij} + \lambda \R_{ijkl} \R^{ijkl} = 2 (d-2) \left[ (d-3) \frac{1-f}{r^2} \left( 1 + \lambda \frac{1-f}{r^2} \right) - \frac{f'}{r} \right].
\ee

\noindent
In the following, we shall use the relation above in order to remove the $\left(f'' \right)^2$ term in (\ref{potential}) and further simplify our next calculations. It is not too hard to obtain: 

\bea
V_{\textsf{T}} [f(r)] &=& f(r) \left( \frac{\ell \left( \ell + d - 3 \right)}{r^2} + \frac{\left( d - 2 \right) \left( d - 4 \right) f(r)}{4r^2} + \frac{\left( d - 2 \right) f'(r)}{2r} \right) + \nonumber \\
&+& \lambda\ \frac{f(r)}{r^2} \left[ \left( \frac{2 \ell \left( \ell + d - 3 \right)}{r} + \frac{\left( d - 4 \right) \left( d - 5 \right) f(r)}{r} + \left( d - 4 \right) f'(r) \right) \left( 2 \frac{1 - f(r)}{r} + f'(r) \right) \right. + \nonumber \\
&+& \left. \Big( 4 (d-3) - (5d-16) f(r) \Big) \frac{f'(r)}{r} - 4 \left( f'(r) \right)^2 + \left( d-4 \right) f(r) f''(r) \right].
\eea

\noindent
This expression is valid for \textit{any} string theory corrected, spherically symmetric, static solution, which has \textit{no} dilaton field at the classical level (as is the case of the CMP solution). The first thing one notices is that the classical term in the potential above, at order ${\mathcal{O}} (\lambda^0)$, precisely matches the IK potential for tensor--type gravitational perturbations, as expressed in (\ref{ikp}). This is certainly to be expected, as the IK potential was derived in pure EH gravity, where there is no dilaton. This also tells us that the CMP solution is, in this context, a very natural stringy extension of the Schwarzschild black hole. We shall use the potential above in the following studies of the CMP geometry.

\subsection{Proof of Perturbative Stability}

We are now fully set to prove stability of the CMP solution. As explained in section 3, stability follows if and only if $Q(r) \ge 0$ for $r \ge R_H$. Using (\ref{f''}) one may re--write $Q$ in (\ref{fpqnp}) as

\bea
Q(r) &=& \frac{f(r)}{1 - 2 \lambda \frac{f'(r)}{r}} \left[ \frac{\ell \left( \ell + d - 3 \right)}{r^2} + 4 \lambda \ell \left( \ell + d - 3 \right) \frac{1-f(r)}{r^4} \right] \nonumber \\
&\simeq& f(r)\ \frac{\ell \left( \ell + d - 3 \right)}{r^2} \left[ 1 + \frac{2\lambda}{r} \left( 2 \frac{1-f(r)}{r} + f'(r) \right) \right].
\eea

\noindent
Because $f(r)>0$ for $r>R_H$ (this is, in some sense, the definition of the event horizon\footnote{Keep in mind that in the present context the CMP solution satisfies $f(r)>0$ for $r>R_H$ as long as the black hole in consideration is large, \textit{i.e.}, as long as one remains in the domain of validity of string perturbation theory.}), it is clear that one will have $Q(r) \ge 0$ for $r \ge R_H$, in any spacetime dimension, as long as 

\be
2 \frac{1-f(r)}{r} + f'(r) > 0
\ee

\noindent
in the very same region. But this term is the stringy correction, so that the function $f(r)$ in it is only to be evaluated at the classical level, \textit{i.e.}, at order ${\mathcal{O}} (\lambda^0)$. In this case it trivially follows for the CMP solution

\be
\left. 2 \frac{1-f(r)}{r} + f'(r) \right|_{{\mathcal{O}} (\lambda^0)} = \frac{2 \MM (d-1)}{r^{d-2}},
\ee

\noindent
which is positive for any $r > R_H$. This proves stability of the CMP black hole solution in any spacetime dimension $d>4$ (where tensor type gravitational perturbations exist), for positive values of the black hole mass and string length. But this result also proves a bit more: all we used above was the generic form of the potential and the classical result for $f(r)$. So, one further concludes that \textit{any} spherically symmetric, static solution, which has no dilaton at order ${\mathcal{O}} (\lambda^0)$, is stable under tensor--type gravitational perturbations.

\section{Scattering Theory in the Callan--Myers--Perry Geometry}

The equation describing gravitational perturbations to the CMP solution is also the equation which allows for a study of scattering in this spacetime geometry. More precisely, it is the equation which allows for a study of greybody factors and quasinormal frequencies, required data in the study of the Hawking emission spectra and quasinormal ringing. In the following, we shall address both these questions.

\subsection{Absorption Cross--Section}

Let us begin with the computation of the CMP black--hole absorption cross--section for low--frequency tensor--type gravitational waves. Starting with \cite{u76}, there is a great deal of literature on greybody factors and black hole absorption cross--sections, computed at low energies, and we refer the reader to \cite{hns06} for a review and complete list of references. A classical result in EH gravity is the fact that, for \textit{any} spherically symmetric black hole in arbitrary dimension, minimally--coupled massless scalar fields have an absorption cross--section which is equal to the area of the black hole horizon \cite{dgm96}. This result is quite spectacular as it points towards a universality of the low--frequency absorption cross--sections of generic black holes in EH gravity. In spite of this, not much work has been done on trying to extend such result away from the EH realm, with the inclusion of higher--derivative corrections. The only exception we are aware of is \cite{gbk05}, where the computation of greybody factors in curvature--squared Lovelock gravity---without a dilaton field---is addressed; which is a different context from the one in this paper. Furthermore, the authors focused more on obtaining numerical results for a wide range of frequencies, rather than on the analytical solution to the low--frequency problem. Here, we shall address the analytical computation of the low--frequency absorption cross--section for the CMP black hole, following the standard analysis described in \cite{u76, hns06}.

As we have said, we will consider scattering of tensor--type gravitational waves, at low frequencies, $R_H \omega \ll 1$. The low--frequency requirement is necessary as we shall use the technique of matching solutions (see \cite{u76, hns06} for details); it is precisely when the wave--length of the scattered field is very large, as compared to the radius of the black hole, that one can actually match solutions near the event horizon to solutions at asymptotic infinity \cite{u76, hns06}. Also, we shall only focus on the leading contribution to the scattering process, given by the s--wave, where $\ell=0$\footnote{At first this might seem puzzling as, for tensor--type perturbations, one should consider $\ell \ge 2$. Our point of view in here is to analyze the simplest possible example, and as such one considers the $\ell=0$ case as a good first approximation---in a partial--wave expansion it is certainly a leading term with respect to all other $\ell>0$ terms. Naturally, future work should consider the case for generic $\ell$. In spite of this, and as we shall find at the end of this section, the result we obtain turns out to be valid \textit{both} as the first approximation to tensor--type gravitational perturbations \textit{and} as the exact result for minimally--coupled massless scalar fields, thus also validating the approximation we choose to make in here, of setting $\ell=0$.}. Let us begin near the CMP black hole event--horizon. At the precise location of the horizon, the potential describing tensor--type gravitational perturbations vanishes, and the master equation reduces to a simple free--field equation whose solutions are either incoming or outgoing plane--waves, in the tortoise coordinate. Very close to the event horizon, $r \simeq R_H$, one has

\bea \label{expa}
V_{\textsf{T}} (r) &\simeq& \frac{(d-2) (d-3)^2}{2} \left( 1 - \frac{(d-1) (d-4)^2}{d-2}\ \frac{\lambda}{R_H^2} \right) \frac{r-R_H}{R_H^3} + {\mathcal{O}} \left( \left( r-R_H \right)^2 \right), \nonumber \\
x (r) &\simeq& \frac{R_H}{d-3} \left( 1 + \frac{(d-1) (d-4)}{2}\ \frac{\lambda}{R_H^2} \right) \log \left( \frac{r-R_H}{R_H} \right) + {\mathcal{O}} \left( r-R_H \right).
\eea

\noindent
The above equations tell us that as long as $\frac{r-R_H}{R_H} \ll \left( R_H \omega \right)^2$ one will have $V_{\textsf{T}} (r) \ll \omega^2$ and in this near--horizon region one may neglect the potential $V_{\textsf{T}} (r)$ in the master equation. One thus obtains, very close to the event horizon,

\be
\left( \frac{d^2}{dx^2} +\omega^2 \right) \Big( k(r) H_T (r) \Big) = 0.
\ee

\noindent
The solutions to the above equation are plane--waves. As we are interested in studying the absorption cross--section, we shall consider the general solution for a purely incoming plane--wave

\be \label{near}
H_T (x) = A_{\text{\tiny{near}}} e^{i \omega x},
\ee

\noindent
where we have also evaluated

\be
k(r) \simeq i R_H^{\frac{d+1}{2}} \left( 1 + \frac{(d+1) (d-4)}{4}\ \frac{\lambda}{R_H^2} \right) + {\mathcal{O}} \left( r-R_H \right),
\ee

\noindent
which can be treated as a constant for $r \simeq R_H$. One may still move slightly away from the event horizon, making use of the above expansion for the tortoise coordinate, (\ref{expa}). Indeed, if one uses (\ref{expa}) in (\ref{near}), one may give one step further out from the black hole event--horizon, while still maintaining the validity of the $\frac{r-R_H}{R_H}$ series--expansion \cite{u76, hns06}. One obtains in this case

\be \label{close}
H_T (r) \simeq A_{\text{\tiny{near}}} \left( 1 + i\frac{R_H \omega}{d-3} \left( 1 + \frac{(d-1) (d-4)}{2}\ \frac{\lambda}{R_H^2} \right) \log \left( \frac{r-R_H}{R_H} \right) \right).
\ee

Another region of spacetime which is simple to study is asymptotic infinity. The asymptotic region of the CMP black hole is precisely the same as the asymptotic region of the Schwarzschild spacetime, which is basically flat Minkowski spacetime. At asymptotic infinity, the potential describing tensor--type gravitational perturbations vanishes, and the master equation reduces to a simple free--field equation whose solutions are either incoming or outgoing plane--waves, in the tortoise coordinate. One may also solve the master equation in the original radial coordinate in terms of Bessel functions, obtaining $H_T (r) = \left( r \omega \right)^{(3-d)/2} \left[ A\, J_{(d-3)/2} \left( r\omega \right) + B\, N_{(d-3)/2} \left( r\omega \right) \right]$ (see, \textit{e.g.}, \cite{dgm96, u76, hns06}). At low--frequencies, with $r\omega \ll 1$, such solution becomes

\be \label{far}
H_T (r) \simeq A_{\text{\tiny{asymp}}}\ \frac{1}{2^{\frac{d-3}{2}} \Gamma \left( \frac{d-1}{2} \right)} + B_{\text{\tiny{asymp}}}\ \frac{2^{\frac{d-3}{2}} \Gamma \left( \frac{d-3}{2} \right)}{\pi \left( r\omega \right)^{d-3}} + {\mathcal{O}} \left( r\omega \right).
\ee

\noindent
In order to compute the absorption cross--section, one now needs to relate the coefficients $A_{\text{\tiny{near}}}$, $A_{\text{\tiny{asymp}}}$ and $B_{\text{\tiny{asymp}}}$. This can be done via a standard technique to match near--horizon to asymptotic solutions, and requires studying the master equation in an intermediate region, between the event horizon and asymptotic infinity (see \cite{u76, hns06} for a clear explanation of this procedure). This is what we shall do in the following.

As one moves away from the black hole event horizon, the potential begins to grow. Because we are studying low--frequency scattering, eventually the potential will be much bigger than the scattering frequency. This is the definition of the intermediate region: it is the region where $V_{\textsf{T}} (r) \gg \omega^2$, but where the low--frequency constraint of $r\omega \ll 1$ remains valid. This constraint, together with $\frac{r-R_H}{R_H} \gg (R_H \omega)^2$, thus defines the radial coordinate range associated to the intermediate region. To proceed, we need to solve the master equation in this region. Let us do this perturbatively in $\lambda$, defining the expansion $H_T (r) = H_0 (r) + \lambda H_1 (r)$. It is not too hard to see that $H_0 (r)$ satisfies

\be \label{h0}
\left[ - f(r) \frac{d}{dr} \left( f(r) \frac{d}{dr} \right) + f(r) \left( \frac{(d-2) (d-4) f(r)}{4 r^2} + \frac{(d-2) f'(r)}{2r} \right) \right] \Big( k(r) H_0 (r) \Big) = 0,
\ee

\noindent
where in here both $k(r)$ and $f(r)$ are to be evaluated at the classical level, \textit{i.e.}, at order ${\mathcal{O}} (\lambda^0)$. Let us define the natural expansions $k (r) = k_0 (r) + \lambda k_1 (r)$ and $f (r) = f_0 (r) + \lambda f_1 (r)$, where the definitions of all terms should be obvious from context. In this case $k_0 (r) = i R_H^{3/2} r^{\frac{d-2}{2}}$ and the previous equation may be re--written as

\be \label{h0simp}
- i R_H^{3/2} r^{\frac{2-d}{2}} f_0(r) \frac{d}{dr} \left( r^{d-2} f_0(r) \frac{d}{dr} \right) H_0 (r) = 0 \quad \Leftrightarrow \quad
\frac{d}{dr} \left( r^{d-2} f_0(r) \frac{d}{dr} \right) H_0 (r) = 0,
\ee

\noindent
whose most general solution is

\be \label{h0sol}
H_0 (r) = A_{\text{\tiny{inter}}}^0 + B_{\text{\tiny{inter}}}^0 \log \left( 1 - \frac{R_H^{d-3}}{r^{d-3}} \right).
\ee

\noindent
Solving for $H_1 (r)$ requires a considerable greater amount of work. One first writes down the equation that $H_1 (r)$ satisfies, and observes that it is a non--homogeneous version of the differential equation for $H_0 (r)$:

\bea \label{h1}
&&
\left[ - f_0 (r) \frac{d}{dr} \left( f_0 (r) \frac{d}{dr} \right) + f_0 (r) \left( \frac{(d-2) (d-4) f_0 (r)}{4 r^2} + \frac{(d-2) f'_0 (r)}{2r} \right) \right] \Big( k_0 (r) H_1 (r) \Big) + \nonumber \\
&+&
\left[ - f_0 (r) \frac{d}{dr} \left( f_0 (r) \frac{d}{dr} \right) + f_0 (r) \left( \frac{(d-2) (d-4) f_0 (r)}{4 r^2} + \frac{(d-2) f'_0 (r)}{2r} \right) \right] \Big( k_1 (r) H_0 (r) \Big) + \nonumber \\
&+&
\left[ - f_0 (r) \frac{d}{dr} \left( f_1 (r) \frac{d}{dr} \right) - f_1 (r) \frac{d}{dr} \left( f_0 (r) \frac{d}{dr} \right) + f_0 (r) (d-2) \left( \frac{(d-4) f_1 (r)}{4 r^2} + \frac{f'_1 (r)}{2r} \right) \right. + \nonumber \\
&+&
\left. f_1 (r) (d-2) \left( \frac{(d-4) f_0 (r)}{4 r^2} + \frac{f'_0 (r)}{2r} \right) + \frac{f_0 (r)}{r^2} \left( (d-4) \left( \frac{(d-5) f_0 (r)}{r} + f'_0 (r) \right) \left( 2 \frac{1-f_0 (r)}{r} + f'_0 (r) \right) + \right. \right. \nonumber \\
&&
\left. \left. + \Big( 4(d-3)-(5d-16) f_0 (r) \Big) \frac{f'_0 (r)}{r} - 4 \left( f'_0 (r) \right)^2 + (d-4) f_0 (r) f''_0 (r) \right) \right] \Big( k_0 (r) H_0 (r) \Big) = 0.
\eea

\noindent
One observes that the first line of the equation above is exactly the same as the differential equation (\ref{h0}), for $H_0 (r)$. We will naturally refer to this line of the differential equation as its ``homogeneous'' part; the only one which depends on $H_1 (r)$. It is not too hard to see that, due to the re--writing (\ref{h0simp}), this is indeed a homogeneous term in the differential equation. The full solution to $H_1 (r)$ is thus given by the general solution to the homogeneous differential equation plus a particular solution to the full non--homogeneous equation (including the terms which do not depend on $H_1 (r)$ in the case above). The homogeneous part of $H_1 (r)$ is simple: it is exactly the same as $H_0 (r)$ in (\ref{h0sol}). The difficult part is, clearly, dealing with the non--homogeneous part of the differential equation. Obtaining an exact solution for the non--homogeneous part of $H_1 (r)$ is a difficult task and beyond the scope of our present analysis. Certainly, this is a problem which should be addressed in future work, in order to better validate the approximations we shall perform in the following. Given what we just said, let us write down the most general solution to this differential equation as

\be
H_1 (r) = A_{\text{\tiny{inter}}}^1 + B_{\text{\tiny{inter}}}^1 \log \left( 1 - \frac{R_H^{d-3}}{r^{d-3}} \right) + H_1^{\text{\tiny{NH}}} (r),
\ee

\noindent
where we do not know the exact form of the non--homogeneous part, $H_1^{\text{\tiny{NH}}} (r)$. For the purpose of this paper, we will proceed by analyzing $H_1^{\text{\tiny{NH}}} (r)$ near the event horizon and near asymptotic infinity. Near the black hole event horizon, where $\frac{r-R_H}{R_H} \ll 1$, the differential equation (\ref{h1}) for $H_1 (r)$ can be approximately written---when expanded to quadratic order in the $\frac{r-R_H}{R_H}$ parameter---as

\bea
&&
- i R_H^{3/2} r^{\frac{2-d}{2}} \left( (d-3) \frac{r-R_H}{R_H} \right) \frac{d}{dr} \left( r^{d-2} \left( (d-3) \frac{r-R_H}{R_H} \right) \frac{d}{dr} \right) H_1^{\text{\tiny{NH}}} (r) + \nonumber \\
&+&
\frac{i}{2} (d-1) (d-3)^2 (d-4) R_H^{\frac{d-7}{2}} \left\{ \left[ - (d-6) B^0_{\text{\tiny{inter}}} + 4 \left( A^0_{\text{\tiny{inter}}} + B^0_{\text{\tiny{inter}}} \log \left( \frac{r-R_H}{R_H} \right) + \right. \right. \right. \nonumber \\
&+&
B^0_{\text{\tiny{inter}}} \log \left( d-3 \right) \bigg) \bigg]\ \left( \frac{r-R_H}{R_H} \right) + \frac{2}{3} \bigg[ \big( (3d-25) d + 26 \big) B^0_{\text{\tiny{inter}}} -12 d\ \bigg( A^0_{\text{\tiny{inter}}} + \nonumber \\
&+&
\left. \left. \left. B^0_{\text{\tiny{inter}}} \log \left( \frac{r-R_H}{R_H} \right) + B^0_{\text{\tiny{inter}}} \log \left( d-3 \right) \right) \right] \left( \frac{r-R_H}{R_H} \right)^2 \right\} = 0,
\eea

\noindent
where we have omitted a long series of manipulations, only keeping in here, for the record, the perturbative result for $k(r)$

\be
k(r) = i R_H^{3/2} r^{\frac{d-2}{2}} \left( 1 + \left( d-4 \right) \left( \frac{d-3}{4} + \frac{R_H^{d-1}}{r^{d-1}} \right) \frac{\lambda}{R_H^2} \right).
\ee

\noindent
A further long calculation finally yields a particular solution to the above non--homogeneous differential equation, expanded in the $\frac{r-R_H}{R_H}$ parameter,

\bea \label{nhnh}
H_1^{\text{\tiny{NH}}} (r) &\simeq& \frac{(d-1)(d-4)}{2 R_H^2} \left[ - (d+2) B^0_{\text{\tiny{inter}}} + 4 \left( A^0_{\text{\tiny{inter}}} + B^0_{\text{\tiny{inter}}} \log \left( \frac{r-R_H}{R_H} \right) + B^0_{\text{\tiny{inter}}} \log (d-3) \right) \right]\ \frac{r-R_H}{R_H} + \nonumber \\
&+& {\mathcal{O}} \left( \left( \frac{r-R_H}{R_H} \right)^2 \log \left(\frac{r-R_H}{R_H}\right) \right),
\eea

\noindent
approximately valid near the black hole horizon, where $\frac{r-R_H}{R_H} \ll 1$. Next, we can use an analogous line of thought near asymptotic infinity. In this case, the differential equation (\ref{h1}) for $H_1 (r)$ can be approximately written to leading order at infinity as

\be
- i R_H^{3/2} r^{\frac{2-d}{2}} \frac{d}{dr} \left( r^{d-2} \frac{d}{dr} \right) H_1^{\text{\tiny{NH}}} (r) - \frac{i}{2} B^0_{\text{\tiny{inter}}} (d-3)^3 (d-4) R_H^{\frac{4d-13}{2}} r^{\frac{6-3d}{2}} = 0,
\ee

\noindent
But the leading term in the potential, which goes as $r^{-\frac{3d}{2}}$, is already sub--leading as compared to the terms in the [homogeneous] differential operator. One thus concludes that, at asymptotic infinity, the terms in the potential are all sub--leading and do not contribute. But then, there is no non--homogeneous term, and the solution at asymptotic infinity is well approximated by the solution to the homogeneous differential equation, which we have obtained earlier. We are now ready to conclude our calculation.

Collecting all relevant formulae above one concludes that, in the intermediate region, the s--wave solution to the master equation is given by

\be \label{interp}
H_T (r) = A_{\text{\tiny{inter}}} + B_{\text{\tiny{inter}}} \log \left( 1 - \frac{R_H^{d-3}}{r^{d-3}} \right) + \lambda H_1^{\text{\tiny{NH}}} (r),
\ee

\noindent
where although we do not know the exact expression for $H_1^{\text{\tiny{NH}}} (r)$, we do know how it approximately behaves near the black hole horizon, as described by (\ref{nhnh}), and we also know that at asymptotic infinity this term can be neglected. It is this expression (\ref{interp}) that allows us to interpolate between the solution near the event horizon of the black hole and the solution at asymptotic infinity, determining relations between several coefficients, and finally leading to the calculation of the absorption cross--section \cite{u76, hns06}. Indeed, near the horizon, $\frac{r-R_H}{R_H} \ll 1$, and

\be
H_T (r) \simeq A_{\text{\tiny{inter}}} + B_{\text{\tiny{inter}}} \log \left( \frac{r-R_H}{R_H} \right) + B_{\text{\tiny{inter}}} \log (d-3) + \cdots.
\ee

\noindent
The first thing one observes is that, in the end, the non--homogeneous contributions turn out to be sub--leading as compared to the homogeneous contributions. In this region one is very close to the black hole event horizon and one may thus match the coefficients above to the ones in (\ref{close}), which immediately yields

\bea
A_{\text{\tiny{near}}} &=& A_{\text{\tiny{inter}}} + B_{\text{\tiny{inter}}} \log (d-3) \simeq A_{\text{\tiny{inter}}}, \nonumber \\
B_{\text{\tiny{inter}}} &=& i A_{\text{\tiny{near}}} \frac{R_H \omega}{d-3} \left( 1 + \frac{(d-1) (d-4)}{2}\ \frac{\lambda}{R_H^2} \right),
\eea

\noindent
where the approximation in the first line follows given both the second line and the low--frequency requirement of $R_H \omega \ll 1$. If instead one evaluates (\ref{interp}) at asymptotic infinity, one finds

\be
H_T (r) \simeq A_{\text{\tiny{inter}}} - B_{\text{\tiny{inter}}} \frac{R_H^{d-3}}{r^{d-3}} + \cdots.
\ee

\noindent
In this region one may match the coefficients above to the ones in (\ref{far}), yielding

\bea
A_{\text{\tiny{asymp}}} &=& 2^{\frac{d-3}{2}} \Gamma \left( \frac{d-1}{2} \right) A_{\text{\tiny{inter}}} = 2^{\frac{d-3}{2}} \Gamma \left( \frac{d-1}{2} \right) A_{\text{\tiny{near}}}, \nonumber \\
B_{\text{\tiny{asymp}}} &=& - \frac{\pi \left( R_H \omega \right)^{d-3}}{2^{\frac{d-3}{2}} \Gamma \left( \frac{d-3}{2} \right)} B_{\text{\tiny{inter}}} = - \frac{i \pi \left( R_H \omega \right)^{d-2}}{2^{\frac{d-3}{2}} (d-3) \Gamma \left( \frac{d-3}{2} \right)} \left( 1 + \frac{(d-1) (d-4)}{2}\ \frac{\lambda}{R_H^2} \right) A_{\text{\tiny{near}}}.
\eea

Computing the low--frequency absorption cross--section is now simple \cite{u76, hns06}. Near the black hole event horizon the total flux is trivial to compute, in the tortoise coordinate, as from (\ref{near}) it immediately follows

\be \label{jnear}
J_{\text{\tiny{near}}} = \frac{A_H}{2i} \left( H_T^\dagger (x) \frac{d H_T}{dx} - H_T (x) \frac{d H_T^\dagger}{dx} \right) = A_H \omega \left| A_{\text{\tiny{near}}} \right|^2.
\ee

\noindent
We want to compare this flux with the incoming flux at infinity. The flux at infinity is computed in the exact same fashion as in the standard EH case, and we refer the reader to \cite{u76, hns06} for details. Here, let us only recall that this flux follows from a free--field equation, as the potential vanishes in this region. As we have alluded to before, in this case one may solve the master equation in the original radial coordinate in terms of Bessel functions. This is quite convenient as it allows for a simple calculation of the flux per unit area associated to an incoming spherical wave, at large $r$---all one has to do is to use the flux formula in (\ref{jnear}), written in the radial coordinate, plus (\ref{far}) for an incoming spherical wave. This flux at infinity is naturally split into incoming and outgoing fluxes as $J_{\text{\tiny{asymp}}} = J_{\text{\tiny{in}}} - J_{\text{\tiny{out}}}$. If we now \textit{impose} the physical requirement of conservation of flux, \textit{i.e.}, that $J_{\text{\tiny{asymp}}} = J_{\text{\tiny{near}}}$, the greybody factor, or absorption probability, in this geometry follows as (see \cite{u76, hns06} for further details)

\be
\gamma(\omega) = \frac{J_{\text{\tiny{near}}}}{J_{\text{\tiny{in}}}} = \frac{J_{\text{\tiny{asymp}}}}{J_{\text{\tiny{in}}}} = 1 - \frac{J_{\text{\tiny{out}}}}{J_{\text{\tiny{in}}}} \simeq 4i\, \frac{B_{\text{\tiny{asymp}}}}{A_{\text{\tiny{asymp}}}}.
\ee

\noindent
The absorption cross--section is now simply obtained via its definition as $\sigma = \gamma |\Psi|^2$, where $|\Psi|^2$ describes the projection of the incoming spherical wave into a plane--wave according to the optical theorem \cite{hns06}. The final result is

\be
\sigma_{\textsf{\tiny{T}}}^{\ell=0} = A_H \left( 1 + \frac{(d-1) (d-4)}{2}\ \frac{\lambda}{R_H^2} \right).
\ee

\noindent
We find that the low--frequency absorption cross--section receives $\a$ corrections with respect to the EH result, although it is \textit{still} proportional to the area of the event horizon. In particular, the cross--section is increased due to the string theoretic corrections. Another observation one can make, following all the steps leading to the result above, is that the ultimate source for the $\a$ corrections in the expression above are the $\a$ corrections present in the tortoise coordinate, (\ref{expa}), \textit{i.e.}, they arise from the stringy modified geometry associated to the expansion $f(r) = f_0 (r) + \lambda f_1 (r)$. In fact, in the expression above, the corrections do not arise from any $\a$ terms in the effective potential or anywhere else.
This fact is quite interesting as it also validates the above analysis for the case of a minimally--coupled massless scalar field. Indeed, as the $\a$ terms in the effective potential end up not contributing to the calculation, the above analysis is exactly the same as if one were to consider \textit{only} the massless scalar field case. The cross--section we obtained thus holds both for a first approximation to tensor--type gravitational perturbations \textit{and} for the exact result of massless scalar fields (in the low--frequency regime).

It is thus not surprising that it is simple to find a more ``geometric'' expression for the absorption cross--section, which may be written as

\be
\sigma_{\textsf{\tiny{T}}}^{\ell=0} = \frac{d-3}{R_H f'(R_H)}\ A_H.
\ee

\noindent
Certainly more evidence would be desirable before establishing this expression as valid to all orders in $\a$. Another important point would be to check whether this expression only holds for uncharged black holes, or if it is of broader validity. We leave such research endeavors for the future.

\subsection{Quasinormal Frequencies}

The study of linear perturbations to black hole solutions yields the following physical picture: after the onset of a perturbation, the return to equilibrium of a black hole spacetime is dominated by damped, single frequency oscillations, which are known as the quasinormal modes (we refer the reader to \cite{n99, ks99} for recent reviews on quasinormal modes and a more complete list of references). These modes are quite special: they depend only on the parameters of the given black hole spacetime, being independent of the details concerning the initial perturbation we started off with. This characteristic quasinormal ringing of black holes, at low frequencies, may be a topic of great future interest as gravitational wave astronomy becomes an experimental reality. Clearly, computing string theoretic corrections---as the ones outlined in this work---to the standard Schwarzschild quasinormal spectrum could be of great interest for a possible astrophysical experimental validation of string theory.

The precise computation of quasinormal frequencies is a hard problem \cite{n99, ks99} and is mostly addressed numerically in the literature. However, in the context of higher--curvature effective actions there is a surprisingly small amount of work done on quasinormal modes, with the notable exceptions of \cite{iiv89, k04, akm05, c06}. These papers have concentrated on the GB non--dilatonic solution of \cite{bd85} and computed quasinormal frequencies of a scalar field in the given background geometry, \textit{i.e.}, solutions of the Klein--Gordon equation in the GB black hole geometry. This work was initiated in \cite{iiv89}, in a four dimensional context, and later extended to $d$--dimensions in \cite{k04}. In \cite{k04, akm05} a thorough numerical calculation of quasinormal frequencies was carried out, including black holes geometries with electric charge or a cosmological constant. As we have alluded to before, although interesting on its own, the non--dilatonic solution of \cite{bd85} is not applicable in a string theory context. Also, if one aims at a future astrophysical experimental validation of these quasinormal frequencies, in any of the upcoming gravitational wave experiments, one should not compute the spectrum of the Klein--Gordon equation but rather the one associated to gravitational perturbations. A first step was given in \cite{c06}, where quasinormal modes of tensor type perturbations to the GB non--dilatonic solution of \cite{bd85}, as described in \cite{dg04, dg05a}, were computed. It would be interesting if the numerical work of \cite{k04, akm05} could be extended to study gravitational perturbations in the present perturbative stringy context.

There has also been some interest in the literature in the computation of quasinormal modes at large imaginary frequencies, the so--called asymptotic quasinormal frequencies. This calculation can actually be done analytically, using a method which involves monodromy matching \cite{mn03} and which is applicable to any static, spherically symmetric black hole solution of general relativity in $d$--dimensions \cite{cns04, ns04}. It would be very interesting to compute string theory corrections to these frequencies. As they are proportional to the black hole temperature on dimensional grounds, it would be quite interesting to know whether the stringy corrections only appear in the $\a$--corrections to the black hole temperature (which we have explicitly displayed earlier), or if they appear in some other different way. Unfortunately, the method of \cite{mn03, ns04} cannot be applied in this situation, as the monodromy matching technique requires full knowledge of the spacetime geometry near the origin \cite{mn03, ns04}: one needs to rotate the Stokes curve, along which the monodromy is computed, around $r=0$ and this requires full knowledge of the function $f(r)$ in the metric in the very same region. While in the Einstein case this is in fact known, in the present context the solution is built perturbatively and is not valid in regions of strong curvature\footnote{There was one attempt to compute asymptotic quasinormal frequencies in the GB non--dilatonic solution of \cite{bd85}, using the monodromy method, in \cite{cg05}. Unfortunately, the authors extrapolated their solution to the region $r\sim0$ where it is not valid.}: indeed $f(r)$ is only valid as long as $r \gg \sqrt{\lambda}$, a fact which is expressed by the statement that the CMP [or any other perturbative] solution is only valid for \textit{large} black holes (\textit{i.e.}, black holes with $R_H \gg \sqrt{\lambda}$, where the horizon is much larger than the string length). There remains the possibility that numerical work will be performed, for one to gain some information on how asymptotic quasinormal frequencies get corrected in the present stringy framework.

\section{Conclusions and Future Directions}

In this paper we have studied static, spherically symmetric black holes in curvature--squared stringy gravity. In particular, we have extended the perturbation theory framework of \cite{ik03a, ik03c} to this string theoretic context and analyzed the stability of the CMP black hole solution under tensor type gravitational perturbations. This allowed us to prove the perturbative stability of the CMP solution, under tensor type perturbations (which were argued in \cite{gh02} to be the only ``dangerous'' modes which could produce instabilities). We have further applied the perturbation theory master equation in the study of greybody factors and computed the absorption cross--section, at low frequencies, for the CMP stringy black hole. We have shown that the low--frequency absorption cross--section is still proportional to the area of the black hole horizon, in spite of receiving $\a$ corrections. The resulting expression for the cross--section can be simply generalized to a formula which could be valid to all orders in $\a$. It would certainly be of great interest to test the validity of our proposal. We have furthermore discussed quasinormal modes in this setting. Many open problems remain to be addressed, and we list a few in the following.

On what concerns perturbative stability, a full proof still requires analysis of vector and scalar type gravitational perturbations. One should proceed along the lines of \cite{ik03a, ik03c}, alongside with the present results, in order to obtain master equations for vector and scalar type perturbations. This would presumably allow for a full proof of stability of the CMP solution. Of course once this is achieved, and because the master equations are valid for any static, spherically symmetric black hole in curvature--squared stringy gravity, in $d$--dimensions, one could also start addressing further solutions, possibly including for dilaton charge, electric charge, and a background cosmological constant.

On what concerns the scattering theory, the first thing to be done is possibly to perform a full numerical investigation of quasinormal frequencies in the present context. This should not only be done at low frequencies, having in mind future gravitational wave experiments, but also at asymptotic frequencies, where the analytical methods of \cite{mn03, ns04} cannot be applied. If one were to obtain further solutions, besides the CMP one, such an analysis should also be carried out for such solutions.

On the theoretical side, one should extend the present results, which concern heterotic as well as bosonic string theory, to the type II supersymmetric case. While this is a problem of great interest, it is also certainly a difficult computational problem. Let us say a few words on this. In \cite{m87} the 
${\mathcal{O}} ({\alpha'}^{3})$ quartic--curvature corrections of type II string theory to the EH action, involving both the graviton and the dilaton, were studied alongside with the analysis of corrections to black hole solutions. String theory corrects the EH and dilaton terms in the supergravity action to (here we remain in the string frame)

\be
S = \frac{1}{2\kappa^{2}} \int \mbox{d}^{10}x \sqrt{-g}\ \mbox{e}^{-2 \phi} \left( \R + 4 \left( \nabla \phi \right)^{2} + \frac{\zeta(3)}{16}\ {\alpha'}^{3}\ Y (\R) \right) + \mbox{fermions},
\ee

\noindent
where $\zeta(s)$ is the Riemann zeta--function, and here

\be
Y (\R) = 2 \R_{\mu\nu\rho\sigma} \R^{\lambda\nu\rho\kappa} {\R^{\mu\alpha\beta}}_{\lambda} {\R_{\kappa\alpha\beta}}^{\sigma} + \R_{\mu\nu\rho\sigma} \R^{\lambda\kappa\rho\sigma} {\R^{\mu\alpha\beta}}_{\lambda} {\R_{\kappa\alpha\beta}}^{\nu}.
\ee

\noindent
It is immediate to realize that the type II supersymmetric case will be much more complicated than the heterotic or bosonic cases which we have studied in the present work. Computing perturbatively in the parameter $\epsilon = \frac{\zeta(3)}{16} {\alpha'}^{3}$, \cite{m87} solved the perturbed field equations to obtain corrections to the $d$--dimensional Schwarzschild solution. The results are qualitatively similar to the ones of heterotic or bosonic theory \cite{m87}: one observes that the dilaton decreases as one approaches the black hole, so that stringy interactions become weaker in the vicinity of the black hole. Moreover, one obtains the result that, for any dimension $d \geq 4$, the black hole temperature is reduced, due to stringy corrections, from the standard Hawking result. This is consistent with the existence of a maximal [Hagedorn] temperature in string theory, and so it is possible that also black hole temperature may reach a maximum value (which then decays to zero, the endpoint of the evaporation being a zero temperature soliton) \cite{cmp89, m87}.

Finally, we should end by noting that black holes in heterotic string theory have lately received a great deal of attention (see, \textit{e.g.}, \cite{cwm98, cwkm00, d04, dkm04, s04}). These (and other) papers showed that starting with certain charged heterotic black holes, which classically have null singularities and vanishing event horizon area, higher--curvature terms actually make these black holes develop a regular horizon, with finite area. It is moreover possible to have a stringy microscopic interpretation of the entropy associated to these derivative--corrected heterotic black holes (although the relation between entropy and horizon area is no longer given by the usual Bekenstein--Hawking formula). It would be quite interesting to develop, for this class of heterotic black holes, similar techniques to the ones we have obtained in this paper. In particular, it would be interesting to compute the absorption cross--section at low--frequencies and check whether it is still proportional to the area of the black hole horizon, in these charged situations. It is possible that the universality of the cross--section may be extended well beyond the EH realm, into full string theory.

\section*{Acknowledgments}
We would like to thank Troels Harmark, Marcos Mari\~no and Jos\'e Nat\'ario for discussions or comments on the draft. The work of FM has been supported by a Chateaubriand scholarship from EGIDE (France) and by the fellowship BPD/14064/2003 from Funda\c c\~ao para a Ci\^encia e a Tecnologia (Portugal).

\vfill

\eject


\bibliographystyle{plain}

\end{document}